\documentclass[aps,prl,twocolumn,superscriptaddress]{revtex4}

\usepackage{graphicx}
\usepackage{amssymb}
\usepackage{pdfpages}

\begin{document}

\title{Structural second-order nonlinearity in metamaterials}

\author{B. Wells}

\affiliation{Department of Physics and Applied Physics, University of Massachusetts Lowell, Lowell, MA, 01854, USA}
\affiliation{Department of Physics, University of Hartford, Hartford, CT}

\author{A.Yu. Bykov}
\affiliation{Department of Physics, King's College London, London, WC2R 2LS, UK}
\author{G. Marino}
\affiliation{Department of Physics, King's College London, London, WC2R 2LS, UK}
\affiliation{Mat\'{e}riaux et Phénoménes Quantiques, Universit\'{e} Paris Diderot-CNRS, F-75013 Paris, France}
\author{M.E. Nasir}
\author{A.V. Zayats}
\affiliation{Department of Physics, King's College London, London, WC2R 2LS, UK}

\author{V.A. Podolskiy}
\email{viktor\_podolskiy@uml.edu}
\affiliation{Department of Physics and Applied Physics, University of Massachusetts Lowell, Lowell, MA, 01854, USA}


\begin{abstract}
Nonlinear processes are at the core of many optical technologies including lasers, information processing, sensing, and security, and require optimised materials suitable for nanoscale integration. Here we demonstrate the emergence of a strong bulk second-order nonlinear response in a composite plasmonic nanorod material comprised of centrosymmetric materials. The metamaterial  provides equally strong generation of the p-polarized second harmonic light in response to both s- and p-polarized excitation. We develop an effective-medium description of the underlying physics, compare its predictions to the experimental results and analyze the limits of its applicability. We show that while the effective medium theory adequately describes the nonlinear polarization, the process of emission of second harmonic light cannot be described in the same framework. The work provides an understanding of the emergent nonlinear optical response in composites and opens a doorway to new nonlinear optical platform designs for integrated nonlinear photonics.
\end{abstract}

\maketitle

A broad class of photonic applications, including frequency conversion, optical information processing, sensing, security, and healthcare, requires materials with second-order nonlinear optical response \cite{BoydBook,shen}. Second harmonic generation, a phenomenon where the incoming radiation of a frequency $\omega$ is converted in the signal at a double frequency $2\omega$, is a fundamental nonlinear process that is used in high-resolution microscopy, optical characterization, and surface studies \cite{SHG-review,SHG-review-2,segovia15}. On the material level, second harmonic generation is described by nonlinear susceptibility tensor, $\hat{\chi}^{(2)}$, which determines the relationship between the excitation fields in the material at the excitation frequency and the induced polarization in the material at the second harmonic frequency \cite{BoydBook,shen}. Due to symmetry considerations, only materials with noncentrosymmetric lattice are capable of strong SHG. While SH signal can be generated in the bulk of centrosymmetric media if one takes into account the higher-order (quadrupolar and magnetic dipolar) terms in the nonlinear polarization expansion, such contributions are generally weak \cite{BoydBook}. Natural optical materials with strong second-order nonlinearity are few, and new solutions are needed to advance nonlinear optics, especially in compact, wavelength-scale and integrated systems. 

Recent advances in nano- and microfabrication have brought into play a new class of composite media, often called metamaterials, where mutual arrangement of the components plays a crucial role in determining their optical properties \cite{pendrySuperlens,enghetaCircuits,kern-17,MiltonBook,PodolskiyBook}. Metamaterials provide a flexible platform for engineering linear optical behavior. Similarly, the effective nonlinear susceptibility of the composite can be related to the nonlinear susceptibilities of the constituent materials. Recently, nonlinear metamaterials have been used for engineering third-order (Kerr-type) nonlinearity, achieving on-demand spectral response, including its sign and polarization control \cite{az-natphot-rev,boydTHG,andres,capolinoENZ,luke}. 

The majority of previous nonlinear metamaterial designs for second-harmonic generation relied on noncentrosymmetric constituents to generate the bulk nonlinear response of the composite  \cite{flytzanisEMT,stroudNLEMT,bergmanNLEMT,hausFWM,boydTHG} and utilized the field enhancement effects to achieve the enhanced nonlinear response. When all constituent materials possess the inversion symmetry, dipolar second-order nonlinear response can only be observed at the interfaces where the symmetry is broken \cite{shennature}. In such  materials SHG originates from a thin surface layer enhanced by the presence of roughness and surface plasmon resonances \cite{sSHG,sipeSurfBulk,shgFortPlasmon,AOM-tomasz,nonl-conr}. 

In this work, we show that the strong bulk nonlinear response of the composites can emerge even if their components possess inversion symmetry. We experimentally demonstrate the second harmonic generation from plasmonic nanorod metamaterials and explain its properties through the effective bulk nonlinearities of the composite. We develop a theoretical description of the observed phenomena and relate the effective volumetric second-order susceptibility of the composite to the material parameters and the arrangement of its components. Finally, we demonstrate that the nonlinear response can be engineered by changing structural parameters of the composite. 

\begin{figure}[b]
\centerline{\includegraphics[width=7.5 cm]{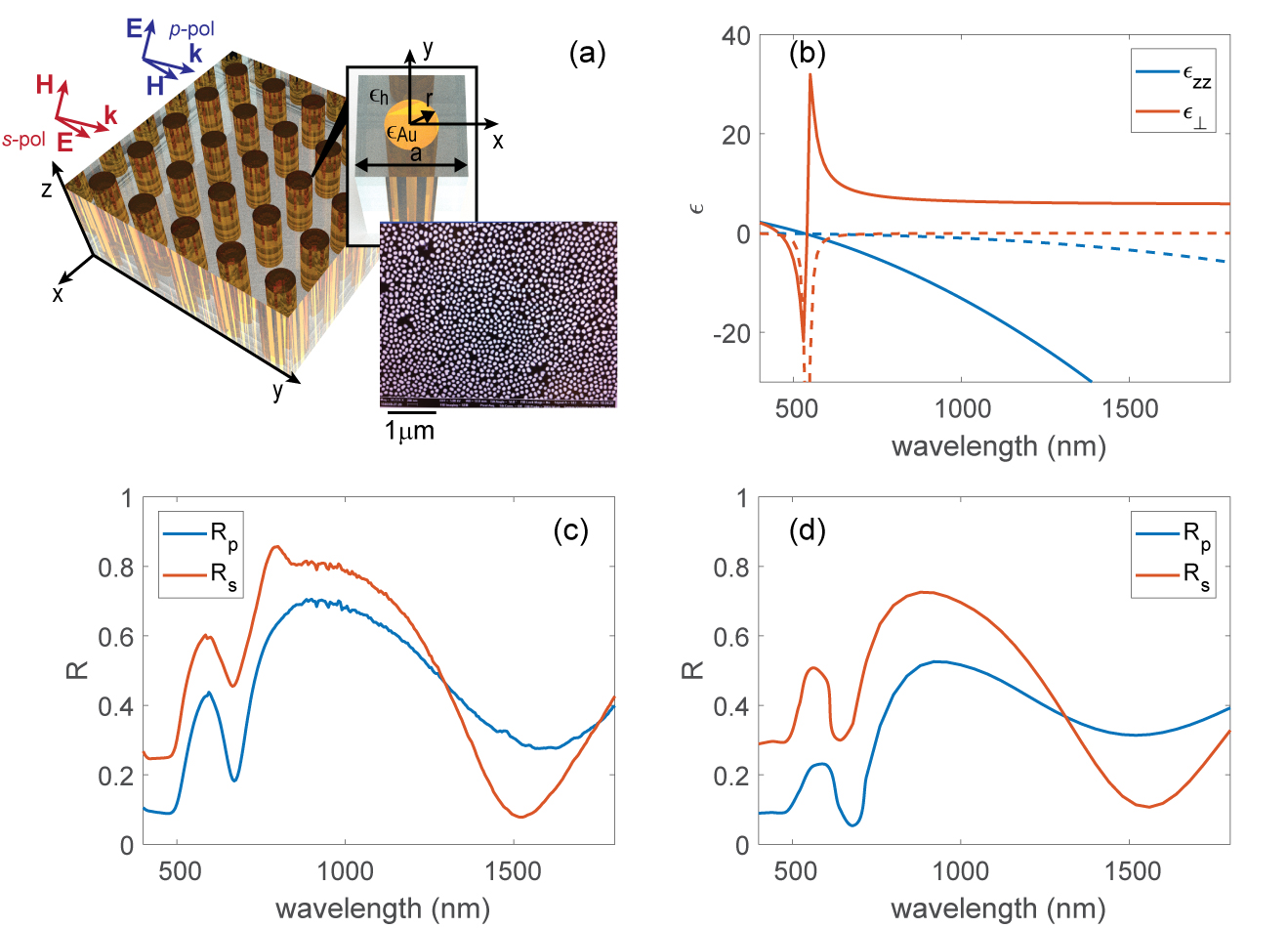}} 
\caption{(color online) 
(a) Schematic of the metamaterial and its unit cell together with the SEM image of the nanorods after the removal
of the AAO matrix. Orientation of the fields and wavevectors considered in modelling and experiments are also shown. (b) Effective medium permittivity of the metamaterial: (solid) real and (dashed) imaginary parts. (c) Measured and (d) simulated reflection spectra of the metamaterial for $p$- and $s$- polarized light at an angle of incidence of 45$^o$. The metamaterial paramaters in (b-d) are $a$=100 nm, $r$=33.5 nm, $h$=150 nm with the nanorods are embedded in the AAO matrix (Sample A).}
 \label{fig1}
\end{figure}

We consider the second harmonic response of the metamaterial comprised of an array of gold nanorods deposited into an alumina matrix (Fig. \ref{fig1}a). When the nanorod radius $r$ and inter-rod separation $a$ are much smaller than the operating wavelength $\lambda$, the metamaterial behaves as an uniaxial crystal with optical axis parallel to the nanorods \cite{podolskiyJOSAB,opex,LPR}. Therefore, its linear optical response is described by a diagonal permittivity tensor $\hat{\epsilon}$ with components $\epsilon_{xx}=\epsilon_{yy}=\epsilon_\perp$ and  $\epsilon_{zz}\neq\epsilon_\perp$. If the material absorption is not too small and the $a\ll\lambda$, the effective medium parameters can be related to the relative permittivity of the host and nanorod materials ($\epsilon_h, \epsilon_{Au}$) as well as the nanorod concentration $p=\pi r^2/a^2$ via
\begin{eqnarray}
\epsilon_\perp=\epsilon_h\frac{(1+p)\epsilon_{Au}+(1-p)\epsilon_h}{(1+p)\epsilon_h+(1-p)\epsilon_{Au}},
\nonumber
\\
\epsilon_{zz}=p\epsilon_{Au}+(1-p)\epsilon_h.
\label{eqEMT}
\end{eqnarray}
The resulting local effective medium theory (EMT) description is known to adequately describe bulk optical response (reflection, transmission, and absorption) of the majority of practical nanorod composites \cite{opex,LPR,nl}. In the limit of small absorption, long nanorods, or large unit cells, the deviations from the local EMT predictions can be quantitatively explained by incorporating the nonlocal (wavevector-dependent) terms into the EMT \cite{podolskiyPRB,footnote}. 

We assume that the relative permittivity of gold is described by the Drude model \cite{johnsonChristy}, which is valid in the spectral range away from the interband transitions: 
\begin{equation}
\label{eqAu}
\epsilon_{Au}= \epsilon_b-\frac{\omega_{pl}^2}{\omega(\omega-i\tau)}
\end{equation}
with the plasma frequency $\omega_{pl}=\sqrt{\frac{e^2 n_0}{m_e \epsilon_0}}=1.36\times 10^{16} s^{-1}$, the inelastic scattering frequency $\tau=1.05\times 10^{14} s^{-1}$, the parameter  $\epsilon_b=9.5$ taking into account background contributions, and $\epsilon_0,e,m_e,n_0$ being the permittivity of free space, the electron charge, the electron mass, and the free-electron density inside gold, respectively. 

Importantly, components of the effective permittivity tensor $\epsilon_\perp$ and $\epsilon_{zz}$ can be of different signs (Fig. \ref{fig1} b), making the nanorod composite a unique “hyperbolic metamaterial” that enables propagation of waveguided modes with subwavelength light confinement that in turn enhance light-matter interaction in the metamaterial \cite{hyperReview1,hyperReview2,PodolskiyBook,LPR,arxiv}.  

From the effective medium standpoint, any monochromatic light propagating in the composite can be represented as a set of plane waves. In particular, when the propagation takes place in $xz$ plane (geometry considered in this work, see insert in Fig. 1), the electromagnetic field can be expressed as a linear combination of $s$- (TE-) and $p$- (TM-) polarized waves. The former have non-zero components of $E_y,H_x,H_z$, while the latter are formed with $E_x,E_z,H_y$ electric and magnetic fields. The effective medium theory adequately represents the $z$-position- and spectral dependence of the unit-cell-averaged fields, as well as the relationship between these averages and the values of the (homogeneous) fields inside the nanorods if the nonlocal effects are not important \cite{SI}.   

Plasmonic nanorod metamaterials were fabricated via Au electrodeposition into nanoporous AAO templates on a glass substrate \cite{fab}. An Al film of 500 nm thickness was deposited on a substrate by magnetron sputtering. The substrate comprises a glass cover slip with a 10-nm-thick adhesive layer of tantalum pentoxide and a 7-nm-thick Au film acting as a weakly conducting layer. Highly ordered, nanoporous AAO was synthesized by a two-step anodization in 0.3M oxalic acid at 40 V. Gold electrodeposition was performed with a three-electrode system using a non-cyanide solution. The length of nanorods was controlled by the electrodeposition time. Fabricated metamaterials were ion-milled to smooth the top surface. The nanorod array parameters used in this work are 150-nm height, 67-nm diameter and 100-nm period (results representing several other samples are provided in SI \cite{SI}). The samples were annealed at 300$^o$C to improve Au optical properties. 

The linear extinction and reflection spectra of the nanorod composit are typical to a hyperbolic metamaterial (Figs. \ref{fig1} and S1 \cite{SI}), showing a minimum due to an overlap of s-polarised ($E_x$) and p-polarised ($E_z$) excited modes and prominent (Fabry-Perot) modes of the metamaterial slab. The measured spectra correspond well to the predictions of both the effective medium theory and the full-wave-numerical simulations \cite{comsol}. The predictions of local effective medium theory are slightly red-shifted with respect to full-wave numerical modelling which can be attributed to nonlocal corrections to the effective medium response \cite{SI}. It is also seen that the effect of the strong absorption by the bound electrons, not included in the Drude model, is significant in the spectral range below 600 nm and can be neglected in the red-near-IR range, which is used in this work.  

Second-harmonic generation spectroscopy was performed using light from the optical parametric amplifier (200 fs pulse trains at the repetition rate of 200 kHz, the average power up to 50 mW in near-IR wavelength range 1100--1800 nm). Note that the metamaterial operates in the hyperbolic regime at both fundamental and second-harmonic frequencies. The laser light polarization was controlled to achieve $p$- or $s$-polarized fundamental light incident on the sample at 45$^o$ with a spot approximately 30--50 $\mu$m in diameter. The reflected $p$- or $s$- polarized second-harmonic light was spectrally selected using the set of short-pass optical filters and measured with the spectrometer and the cooled CCD camera. In order to compensate for pulse energy and pulse duration fluctuations, the measured signal was normalized to a reference SHG measured in reflection from $\beta-$BBO crystal.
   
\begin{figure}
\centerline{\includegraphics[width=7.5 cm]{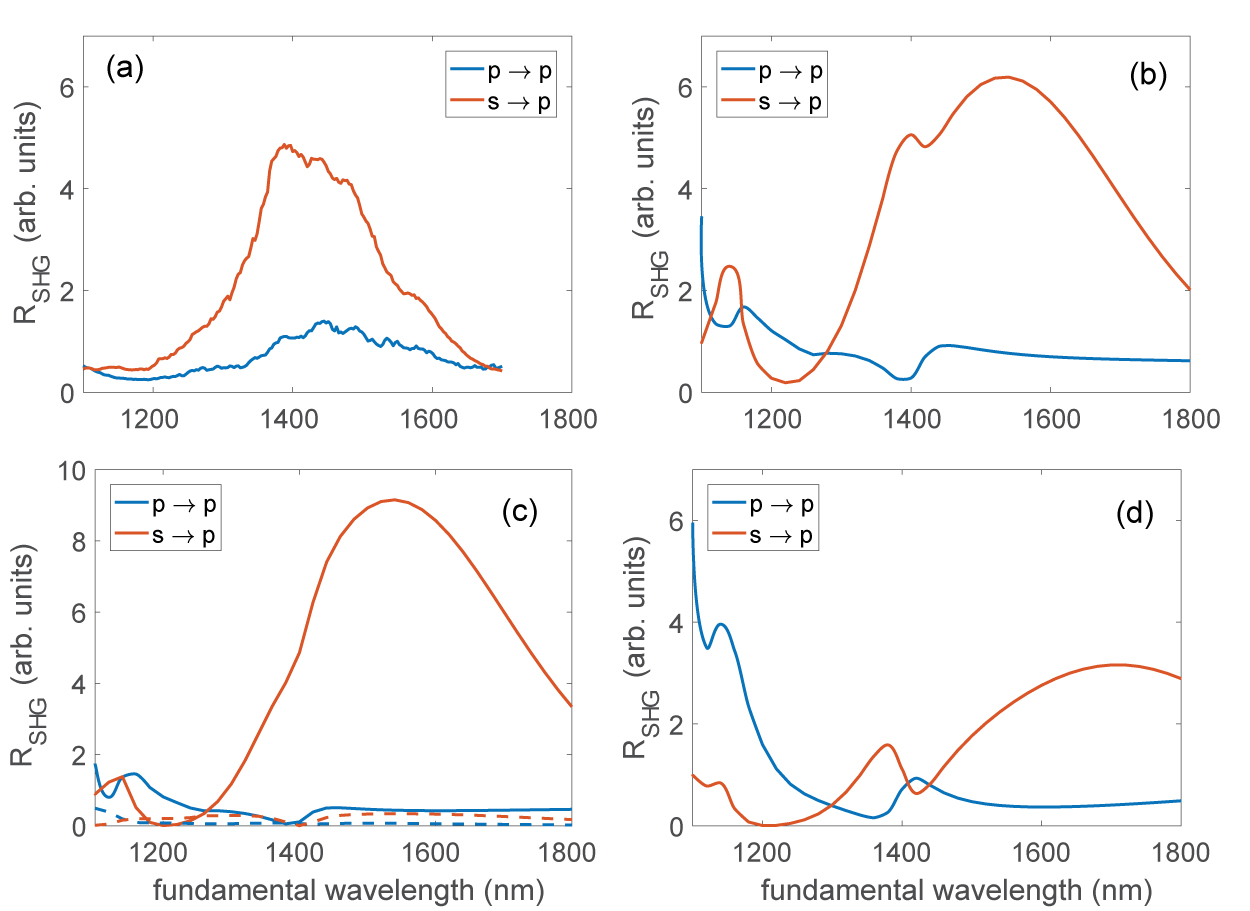}} 
\caption{
(color online) SHG spectra for different polarization configurations from the metamaterial Sample A at an angle of incidence of 45$^o$: (a) experiment, (b) full-wave  numerical simulations, (c) simulations with the nonlinear polarization described by (solid lines) Eq. (S2) \cite{SI} and (dashed lines) only by the additional to Eq. (S2) \cite{SI} components in Eq.(\ref{eqHydro}), (d) SHG spectra simulated with the nonlinear EMT model [Eqs. (\ref{eqQuadForm},\ref{eqChi2})].}
 \label{fig2}
\end{figure}

The SHG spectra measured in the hyperbolic dispersion range (Figs. \ref{fig2}a and S1 \cite{SI}), exhibit pronounced maxima associated with excitation of the metamaterial slab modes \cite{arxiv} for both p- and s-polarised excitation, with the SH light being always p-polarised. The shift of the peak wavelength corresponds to the shift of the mode positions observed for different polarisations in the linear reflection spectra. Interestingly, under s-polarised excitation, SH intensity is approximately 4 times stronger, indicating to the important role of the local fields inside the metamaterial as was observed previously for the nanoparticle composites \cite{AOM-tomasz}.  

The spectral and polarization dependences of the SHG are in a good agreement with the full-wave numerical simulations (Fig. \ref{fig2} b) which implement the hydrodynamic model of the SHG generation in plasmonic media \cite{moloneyHydro,arxiv,al-nat-comm-nonlocal} (see SI \cite{SI} for the details of the numerical simulations). In this model the nonlinear polarization of gold is given by 
\begin{eqnarray}
{\textbf{P}_{2\omega}}=N\left\{\sum_\alpha\frac{\partial}{\partial r_\alpha}\left(\frac{{\bf j_\omega}j_{\omega;\alpha}}{e n_0}\right)
-\right.
\nonumber
\\
\left.
\frac{e}{m_e}[\epsilon_0 ({\bf \nabla}\cdot{\bf E_\omega}){\bf E_\omega}+{\bf j_\omega}\times{\bf B_\omega}]\right\},
\label{eqHydro}
\end{eqnarray}
where $N=(2 \omega(2\omega-i\tau))^{-1}$, $\omega$ and $2\omega$ represent the fundamental and second harmonic frequencies,  index $\alpha$ represents the Cartesian coordinates, and $\bf{E,B,j}$ are the electric field, the magnetic induction, and the current density, respectively \cite{SI}.  

The detailed analysis shows that the SHG efficiency and polarization dependencies are complex functions of the effective medium parameters, thickness of the metamaterial slab, and angle of illumination $\theta$ \cite{arxiv}. Nevertheless, in all cases the nonlinear polarization, and thus SHG generation is dominated by the terms related to the components of the electromagnetic fields that have nonvanishing unit-cell-averages and to $\partial/\partial z$ derivatives of these components (see Eq.(S2) in SI \cite{SI} for explicit expressions). These terms correctly reproduce spectral response of the SHG emission while slightly over-estimating reflected SHG (Fig. 2c). At the same time, Eq.(S2) under-estimates transmitted SHG, so that total SHG intensity calculated from Eq.(S2) is in line with the full numerical solutions of the Maxwell equations. It should be noted that taking into account the SHG from the surface of a metamaterial slab leads to some re-distribution of reflected and transmitted SHG, leaving total SHG intensity practically unchanged (Fig. S2).

It becomes possible to represent the unit-cell-average  nonlinear polarization in the metamaterial (see SI \cite{SI} for derivations) as a quadratic form of the (unit-cell-averaged) fields, essentially introducing effective {\it bulk} second-order nonlinear susceptibilities $\chi^{(2,e)}$ and $\chi^{(2,m)}$   
\begin{equation}
\label{eqQuadForm}
P_{2\omega;\alpha}=\sum_{\beta,\gamma}\left[\chi^{(2,e)}_{\alpha;\beta\gamma}E_{\omega;\beta}E_{\omega;\gamma}+\chi^{(2,m)}_{\alpha;\beta\gamma}E_{\omega;\beta}H_{\omega;\gamma}\right].  
\end{equation}
where the Greek subscripts  represent the Cartesian coordinates $x,y$, and $z$.

The components of the effective nonlinear susceptibility were calculated in the limit of the validity of the local EMT [Eq.(\ref{eqEMT})] that yields homogeneous fields across the crossection of the nanorods \cite{podolskiyJOSAB,SI} by substituting explicit relationships between the field components inside the nanorod, their unit-cell-averages, frequency, and the components of the wavevector, resulting in 
\begin{eqnarray}
\chi^{(2,e)}_{x;xx}=-NL\cdot\frac{i\sigma_\omega^2\epsilon_\perp k_x}{n_0 e\epsilon_{zz}}, 
\nonumber
\\
\chi^{(2,e)}_{x;zz}=-N\cdot\left(\frac{e\sigma_\omega p \omega\epsilon_{zz}}{m_e c^2 k_x}-
            \frac{\sigma_\omega^2(k_x^2-\epsilon_{zz}\omega^2/ c^2)}{n_0 e k_x}L\right), 
\nonumber
\\
\chi^{(2,e)}_{z;xz}=-N/2\cdot\left(\frac{e\sigma_\omega\omega\epsilon_{zz}}{m_e c^2 k_x}L+
	\frac{2 i\sigma_\omega^2 p k_x\epsilon_\perp}{n_0 e\epsilon_{zz}}\right), 
\nonumber
\\
\chi^{(2,e)}_{x;yy}=NL\cdot\frac{k_x  e\sigma_\omega}{\omega m_e},
\nonumber
\\
\chi^{(2,m)}_{z;yx}=-NL\cdot\frac{e\sigma_\omega\mu_0}{m_e}
\label{eqChi2}
\end{eqnarray}
Here,  $k_x=\omega \sin \theta/c$ is the transverse component of the wavevector, $\sigma_\omega=i\omega\epsilon_0\epsilon_{Au}(\omega)$ is the frequency-dependent conductivity of gold, and $L=2p\epsilon_h/[\epsilon_{Au}(\omega)+\epsilon_h]$ represents the relationship between the $E_x,E_y$ components of the electric field inside the nanorod and its unit-cell-averaged values. The first three components describe SHG excitation due to $p$-polarized fundamental light, while the latter two represent the SH generated by the $s$-polarized beam (the second harmonic radiation is entirely $p$-polarized). 

Equations (\ref{eqQuadForm},\ref{eqChi2}) represent the main conclusions of this work: the metamaterial as a whole exhibits dipolar-like nonlinear response even though its material constituents lack bulk dipolar $\chi^{(2)}$. The effective nonlinear susceptibilities are determined by the nonlinear susceptibilities of the constituent materials of the composite and the structure of the local fields inside it \cite{note-re-tensor}. The components of the effective nonlinear susceptibility depend on an angle of incidence so that the symmetry of the metamaterial is broken by the internal fields, except at normal incidence when the electric dipole SHG is forbidden due to symmetry considerations. The explicit dependence of the effective nonlinear susceptibility on the wavenumber reflects the {\it structural} origin of the nonlinearity of a metamaterial. 

\begin{figure}[b]
\centerline{\includegraphics[width=7.5 cm]{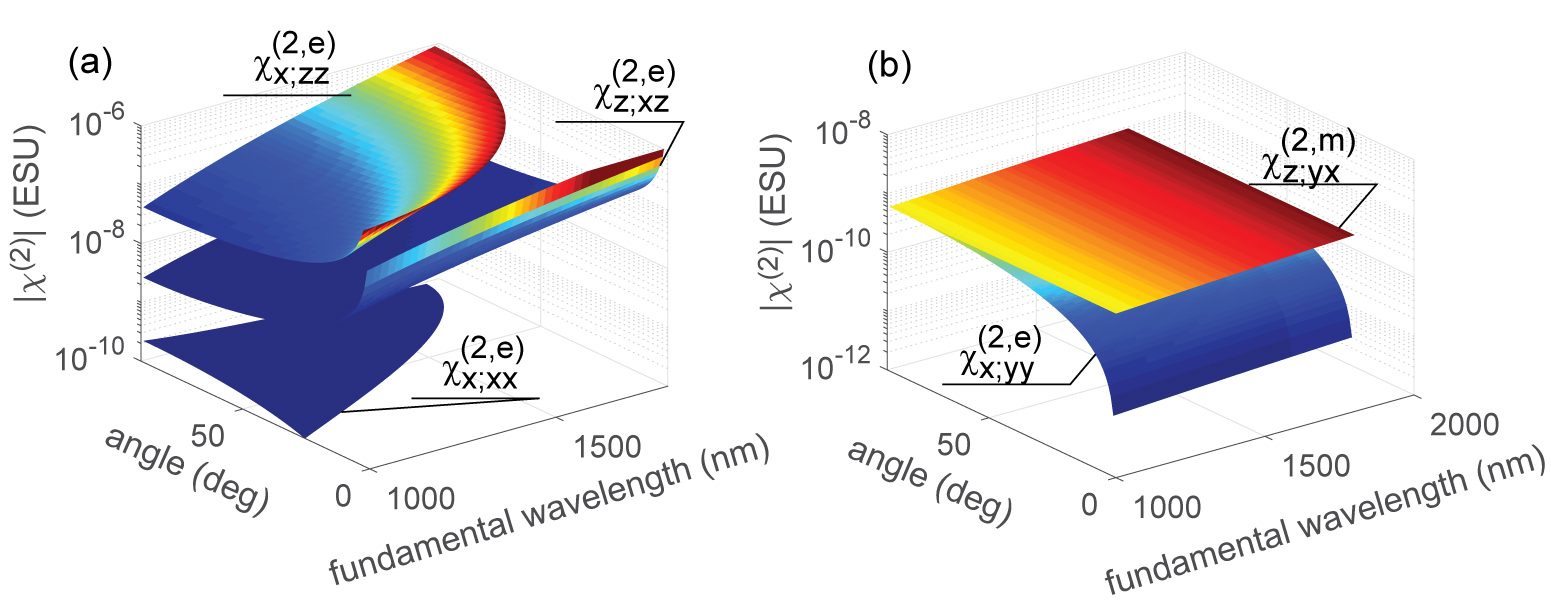}} 
\caption{(color online) Spectral and angular dependences of the components of the effective nonlinear susceptibility [Eq.(\ref{eqChi2})] for (a) $p$-polarized and (b) $s$-polarized fundamental light.}
 \label{fig3}
\end{figure}

The developed nonlinear effective medium theory adequately predicts both spatial distribution and spectral response of the nonlinear polarization in the nanorod composite with exception of small red-shift of the SHG spectra that is likely related to the deviation from the local EMT [Eq.(\ref{eqEMT})] (Figs. \ref{fig2} c, S14 and S15 \cite{SI}). The calculated values of an effective nonlinear response of the composite $\chi^{(2)}\sim10^{-10} - 10^{-7}$ [SGS units] (Fig.\ref{fig3}) indicate relatively strong nonlinearity, comparable to common nonlinear-optical crystals like LiNbO$_3$ and KDP \cite{shen}.

In contrast to common nonlinear optical crystals, the structural origin of the second-order nonlinearity in metamaterials provides a platform for engineering not only wavelength but also polarization properties of a nonlinear response. For example, the structural parameters of the nanorod metamaterials can be tuned to achieve dominant contribution from either $s\rightarrow p$ (Fig. \ref{fig2} a,b) or $p\rightarrow p$ SHG polarization configurations. 

\begin{figure}[t]
\centerline{\includegraphics[width=7.5 cm]{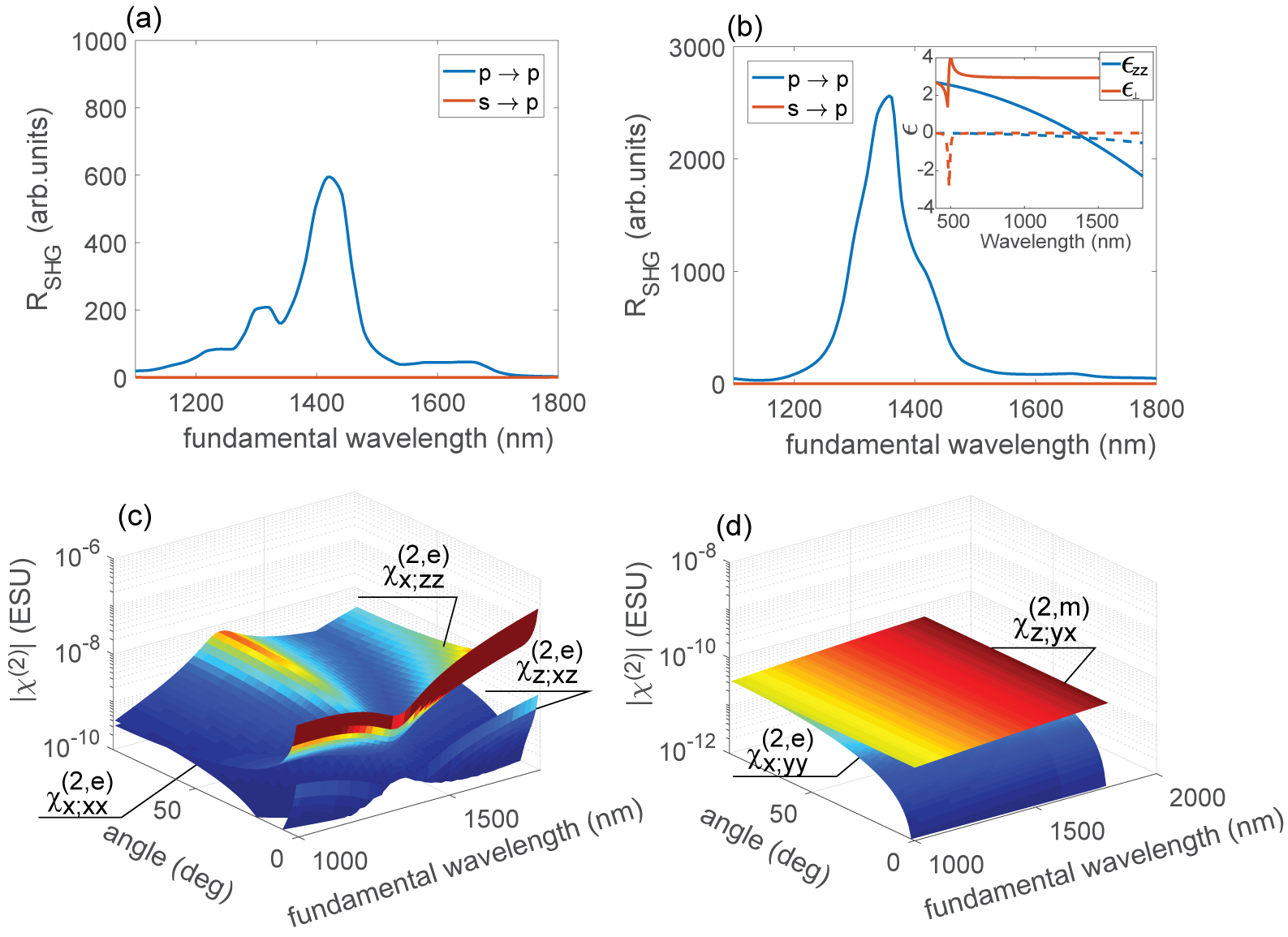}} 
\caption{
(color online) SHG dependences in the ENZ regime (the nanorod metamaterial parameters are h=400 nm, a=100 nm, r=10 nm): (a) full-wave numerical simulations and (b) nonlinear EMT. (c,d) The effective nonlinear susceptibility components for (c) $p$-polarized and (d) $s$-polarized fundamental light.}
 \label{fig4}
\end{figure}

An example of such tuning is shown in Fig. \ref{fig4} that summarizes the SHG simulated from a 400 nm-thick composite with $r=10$ nm, $a=100$ nm. The smaller metal concentration in this composite pushes the effective plasma frequency \cite{LPR} at which $\epsilon_{zz}\simeq 0$ to $\lambda_0\simeq$ 1400 nm, drastically enhancing the $z$-component of electric field inside the composite (see Fig.S8), and thus enhancing $p\rightarrow p$ SHG in the vicinity to the effective plasma frequency (similar response has been predicted for bulk, nontuneable, epsilon-near-zero (ENZ) materials \cite{scaloraENZ,capolinoENZ}). Note that the enhancement of the local field inside the ENZ composite more than compenastes the reduction of the effective $\chi^{(2)}$ (cf. Figs.\ref{fig3} and \ref{fig4}).  

The main limitation on the effective medium nonlinear description, presented in this work, comes from the granularity of metamaterial. In particular, the local EMT that underpins the final expressions for Eq.(\ref{eqEMT}) fails to describe excitation of cylindrical plasmons along the nanorods. While these excitations have limited effect on transmission and reflection from the composite (with possible exception of the elliptic and ENZ ranges), the existence of these excitations drastically affects emission from the composite, as has already been observed for dipolar emitters inside a hyperbolic metamaterial \cite{LSA-purcell}. As the result, the nonlinear effective medium theory, presented in this work, adequately describes distribution of the nonlinear polarization across the composite when the fundamental light frequency is in the hyperbolic dispersion range. The description of the nonlinear polarization distribution becomes less accurate across the elliptic and ENZ frequency ranges (Figs. \ref{fig4}, S8, S13). In any case, the EMT theory cannot be used to predict the emission of light from the metamaterial \cite{SI}. We expect that including the high-index “longitudinal” modes through the nonlocal effective medium theory \cite{podolskiyPRB,LSA-purcell,ak-acs-phot} may address the shortcomings of the formalism presented in this work. Nonlocal response of free-electron plasma \cite{hausNonlocal} may become relevant for composites with drastically thinner wires. 

In conclusion, we have demonstrated the emergence of the structural nonlinearity in composite metamaterials. The approach, presented here on the example of second-harmonic generation from plasmonic nanorod metamaterials, can be extended to analyze nonlinear response of a broad class of composites, such as plasmonic nanoparticle metasurfaces \cite{AOM-tomasz} and metamaterials based on noncentrosymmetric, strongly nonlinear materials, such as AlGaAs nanopillars \cite{algaas}. Structural nonlinearity opens to door to utilize composite media to engineer spectral and polarization nonlinear response beyond what is available with naturally occurring materials.

This work has been supported in part by ARO (grants \# W911NF-12-1-0533, W911NF-16-1-0261), EPSRC (UK), and the ERC iPLASMM
project (321268). A.V.Z. acknowledges support from the Royal Society and the Wolfson
Foundation. The data availability statement: all the data supporting this research are provided in
the article or Supplementary Information.


\begin{thebibliography}{45}
\bibitem{footnote} Note that these nonlocal corrections originate from the composite nature of the plasmonic metamaterial and not from the optical response of its components which is assumed to be described by local $\epsilon_{Au}$ and $\epsilon_h$.
\bibitem{note-re-tensor} We limit our study to the regime when the sample is excited by a single electromagnetic wave that is either $p$- or $s$-polarized. The detailed investigation of more complicated excitation geometries and the analysis of tensorial properties of $\chi^{(2)}_{\alpha;\beta\gamma}$ will be subject of future work.
\end{thebibliography}


\onecolumngrid


\section*{Supplementary material}

# Supplementary Information

## 1. Linear optical properties

Three different nanowire samples, Sample A (150 $nm$ height, $r = 33.5\ nm$, $a = 100\ nm$), Sample B (200 $nm$ height, $r = 27\ nm$, $a = 100\ nm$), and Sample C (390 $nm$ height, $r = 30\ nm$, $a = 100\ nm$) were analyzed in the experiments. In all the samples, gold wires were grown inside alumina matrix. Linear reflection spectra of these samples are shown in Fig. S1. In addition, in the theoretical calculations, we have considered optical response of Sample D (400 $nm$ height, $r = 10\ nm$, $a = 100\ nm$).

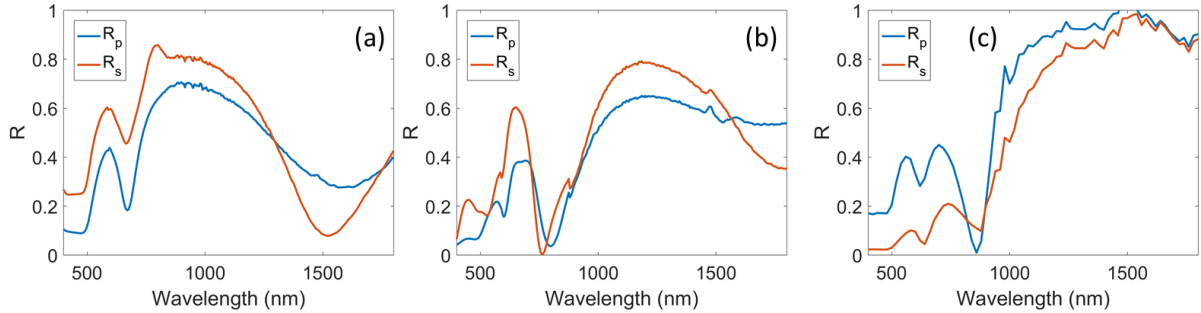

*Fig. S1. Reflection spectra of the samples A (a), B (b), and C (c) for p (blue lines) and s (red lines) polarized plane–wave illumination at an angle of incidence of 45°.*

## 2. Full-wave numerical solutions of Maxwell equations

Maxwell equations have been solved with commercially-implemented finite-element-method (FEM) partial differential equation (PDE) solver [39]. In these calculations, a metamaterial is assumed to be periodic in the $x$ and $y$ directions with a unit cell shown in Fig.1a (inset). We use Floquet-periodicity boundary conditions with $k_x = \frac{\omega}{c}\sin\theta$ *and* $k_y = 0$ for all vertical boundaries of the geometry, with $\omega, c$ and $\theta$ being the angular frequency of light, the speed of light in vacuum, and the angle of incidence (in vacuum), respectively. In the $z$ direction, the geometry is terminated with the COMSOL ports (see documentation [39] for details of FEM implementation of ports and periodic boundary conditions). Our implementation assures "phase-matching" condition $k_x^{(2\omega)} = 2k_x^{(\omega)}$ between fundamental and second harmonic frequencies [1,35].

At a fundamental frequency, a single port is used for the excitation, allowing for calculations of macroscopic properties (transmission and reflection) as well as spatial distributions of electric and magnetic fields across the metamaterial. These distributions are then used to calculate the distribution of the nonlinear polarization (see Section 5 in SI). Our study suggests that discontinuity of normal components of the electric field across the wire interface, accompanied by non-rectangular mesh used by FEM solver, tend to introduce numerical artifacts into the calculations of nonlinear source distribution. To avoid these artifacts, the field distribution calculated by the FEM were re-meshed onto rectangular grid, and all the derivatives were calculated using central-difference scheme [i] using in-house codes.

At a second harmonic frequency, all the input ports of the FEM model were turned off, and the light was generated solely by the nonlinear current term calculated from the distribution of the nonlinear polarization. To mitigate potential numerical artifacts caused by sharp edges of cylindrical nanowires, nonlinear polarization in the top and bottom 15-nm-thick layers of the nanowire composites was zeroed-

out. Our tests (Fig.S2) indicate that total SHG signal is rather robust with respect to edge-related regularization. These tests also indicate the majority of SHG signal originate in the bulk of the manuscript, not on metal-air interface.

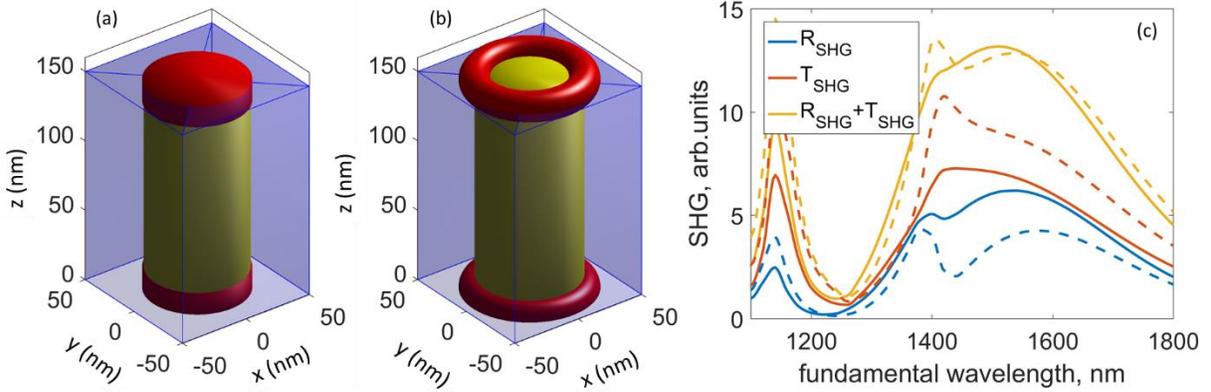

*Fig. S2. (a,b) Unit cells of the nanowire metamaterial with the red areas removed from the numerical simulations in (c) in order to remove numerical artifacts caused by the sharp wire edges. (c) SHG spectra from the sample A (s → p configuration) calculated with the regularization according to geometry of (solid lines) panel (a) and used throughout other simulations and (dashed lines) panel (b). As can be seen, changing regularization technique may slightly re-distribute transmitted and reflected SHG channels while it does not affect total SHG intensity.*

## 3. Nonlinear effective medium simulations

Two additional models were created to verify the validity of the nonlinear effective medium theory described in the main text. The first model combines the simulations at fundamental and second harmonic frequencies that are run sequentially. The effective medium response is assumed at the fundamental frequency, resulting in the unit-cell-averaged field distribution. This field distribution, along with the effective nonlinear polarizability Eqs. (4-5), is then used to directly calculate the distribution of the unit-cell-averaged nonlinear polarization. After rescaling to preserve the unit-cell-average, this nonlinear polarization is then used in the full-3D second harmonic model inside the metamaterial. The last step of the model calculates the emission of the second harmonic light. The results of this model are shown in Figs. 2d, 4d, S14d, and S15d.

Finally, in order to understand the limits of validity of the nonlinear effective medium theory, we have created a model where the composite was replaced by its homogenized metamaterial analog at both fundamental and second harmonic frequencies. Similar to the previous case, the nonlinear polarization was calculated by combining the effective nonlinear polarizability and the (unit-cell-averaged) fields at the fundamental frequency. The results of this model are shown in the insets in Figs.S14d and S15d.

## 4. Linear EMT response, validity or average fields, R/T, effect of gold permittivity

Linear response of metamaterials was analyzed with the full-wave FEM solutions of the Maxwell equations as well as with the transfer matrix formalism[ii,iii]. In the latter case, the response of the nanowire metamaterial has been approximated by the effective medium theory.

Previous studies of optical response of nanorod metamaterials [iv,v,vi] has demonstrated that the permittivity of gold is adequately describes the experimental Johnson-Cristy (JC) data [32], corrected to

take into account the restricted mean free path of electrons in electrochemically-derived gold according to the model from Ref. [vii]. At the same time, the hydrodynamic model of nonlinearity in plasmonic media [40,41] assumes analytical Drude-like response of the permittivity of gold.

Permittivity of gold, calculated according to Refs. [32,vii] and Drude models are compared to each other in Fig. S3. It is seen that the two models deviate only across the visible part of the spectrum. The two models agree across the near-IR frequency range considered in this work.

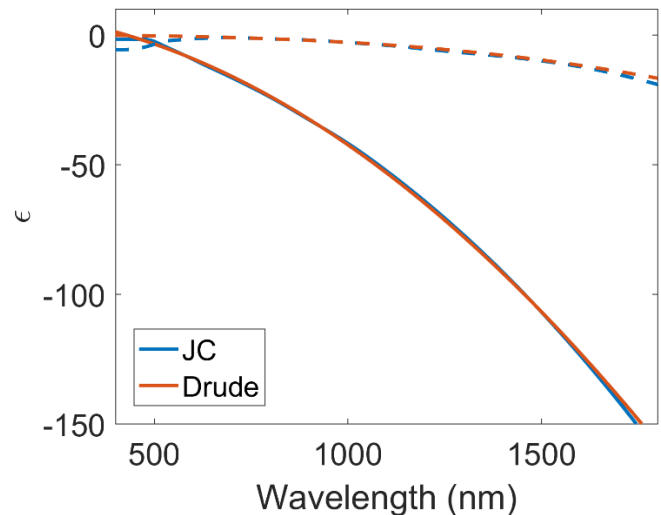

*Fig. S3. Permittivity of gold calculated using Refs. [32,vii] and Drude model; solid and dashed lines represent the real and imaginary parts of the permittivity, respectively.*

The effect of the permittivity model on the optical response of the nanorod metamaterial is illustrated in Fig. S4 that compares the response of the three samples A, B, and C, when calculated with different permittivity models for gold and different approaches to solving the Maxwell equations.

It is seen that throughout the near-IR frequency range the optical response of the composite modeled with Refs. [32,vii] permittivity is similar to that resulting from the idealized Drude response.

The effective medium description yields a small red-shift of the absorption resonances as compared to the full-wave solutions of the Maxwell's equations, an effect likely resulting from limitations of the local effective medium response [viii]. Nevertheless, with exception of the above red-shift, the effective medium theory is seen as a powerful tool for understanding the optics of nanowire composites.

To further assess the validity of the effective medium description of the metamaterials, we have used both the FEM and transfer matrix calculations to plot spectral and spatial profiles of the electromagnetic fields throughout the composites (Figs. S5—S8). Once again, it is seen that the effective medium adequately describes optical behavior of the composites, with exception of the red-shift, already mentioned above, and a small ~20-nm-long regions in the vicinity of the nanowire edges where excitation of the additional wave (primarily seen in $E_z$ field) is noticeable in the FEM calculations (breakdown of the local EMT).

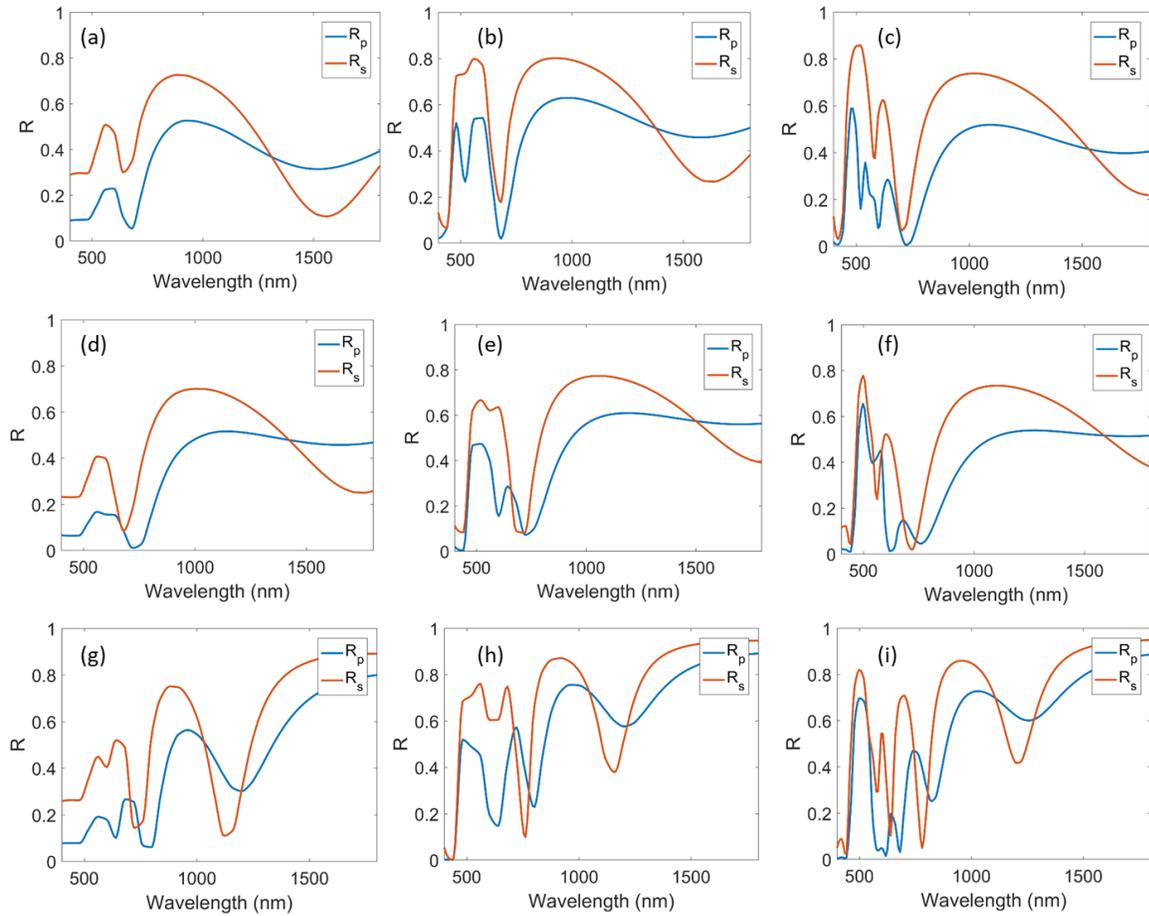

*Fig. S4. Transmission and reflection spectra of the samples A (a,b,c), B (d,e,f), and C (g,h,i), calculated with the full-wave FEM solutions of the Maxwell's equations (a,b,d,f,g,h), and with the transfer matrix formalism (c,f,i).*

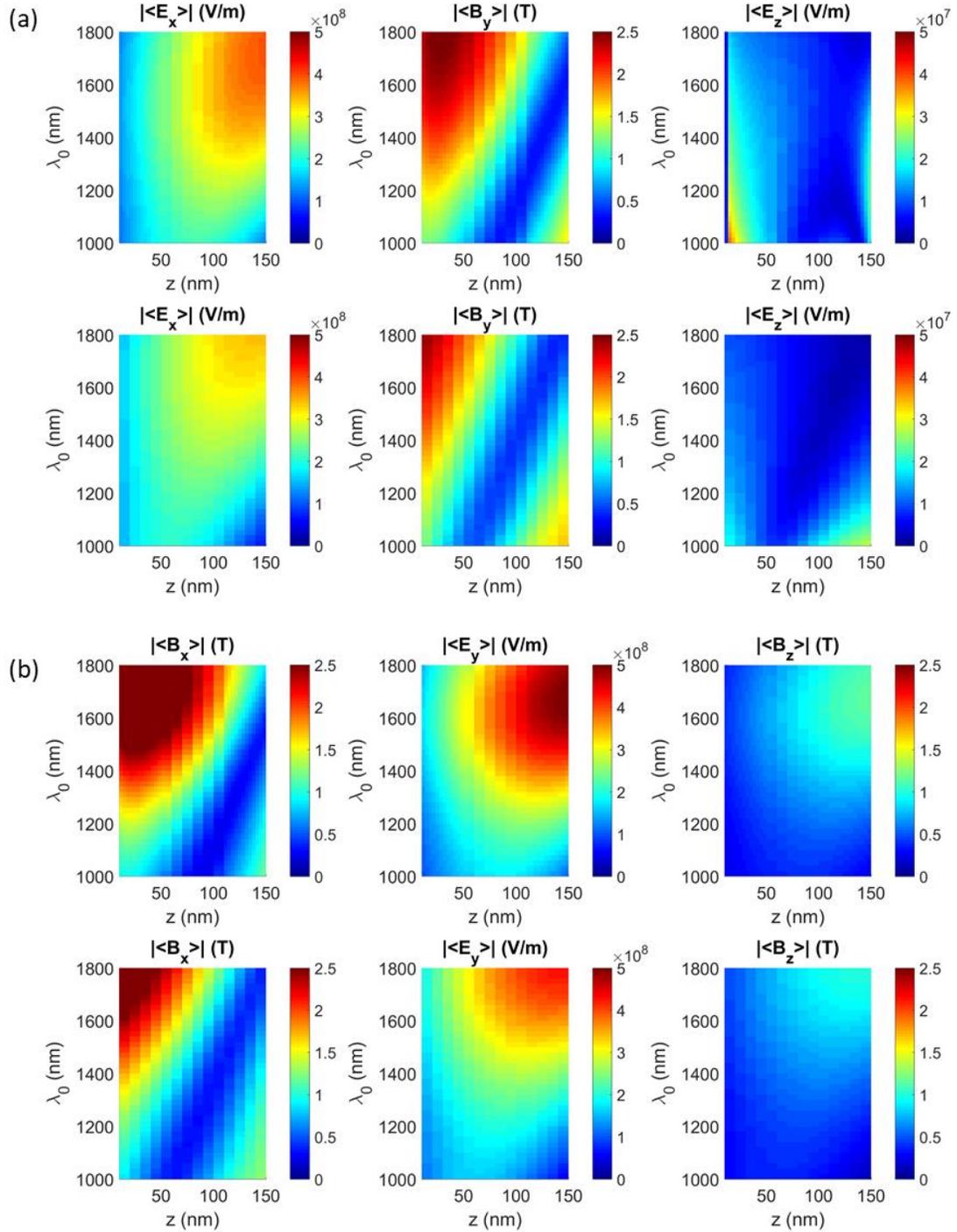

Fig. S5. Comparison between position- and spectral behavior of non-vanishing unit-cell-averaged fields when the Sample A is excited by p-polarized (a) and s-polarized (b) light. Amplitude of the incident field in the FEM calculations are chosen to produce a unit energy flux through a unit cell; the amplitudes of the incident field in the TMM calculations are chosen to match the amplitude of the incident field in the FEM.

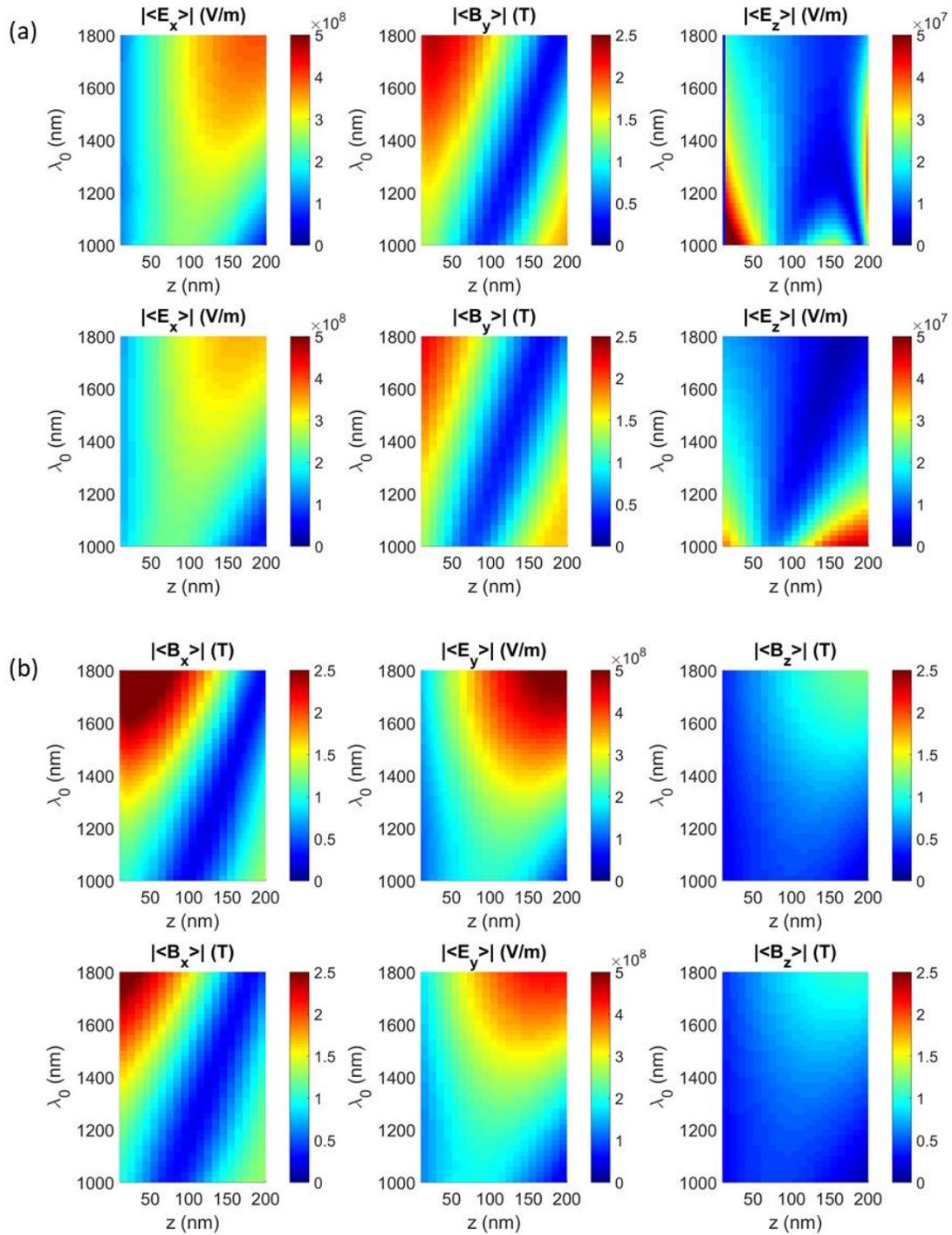

*Fig. S6. Same as Fig. S5 for Sample B.*

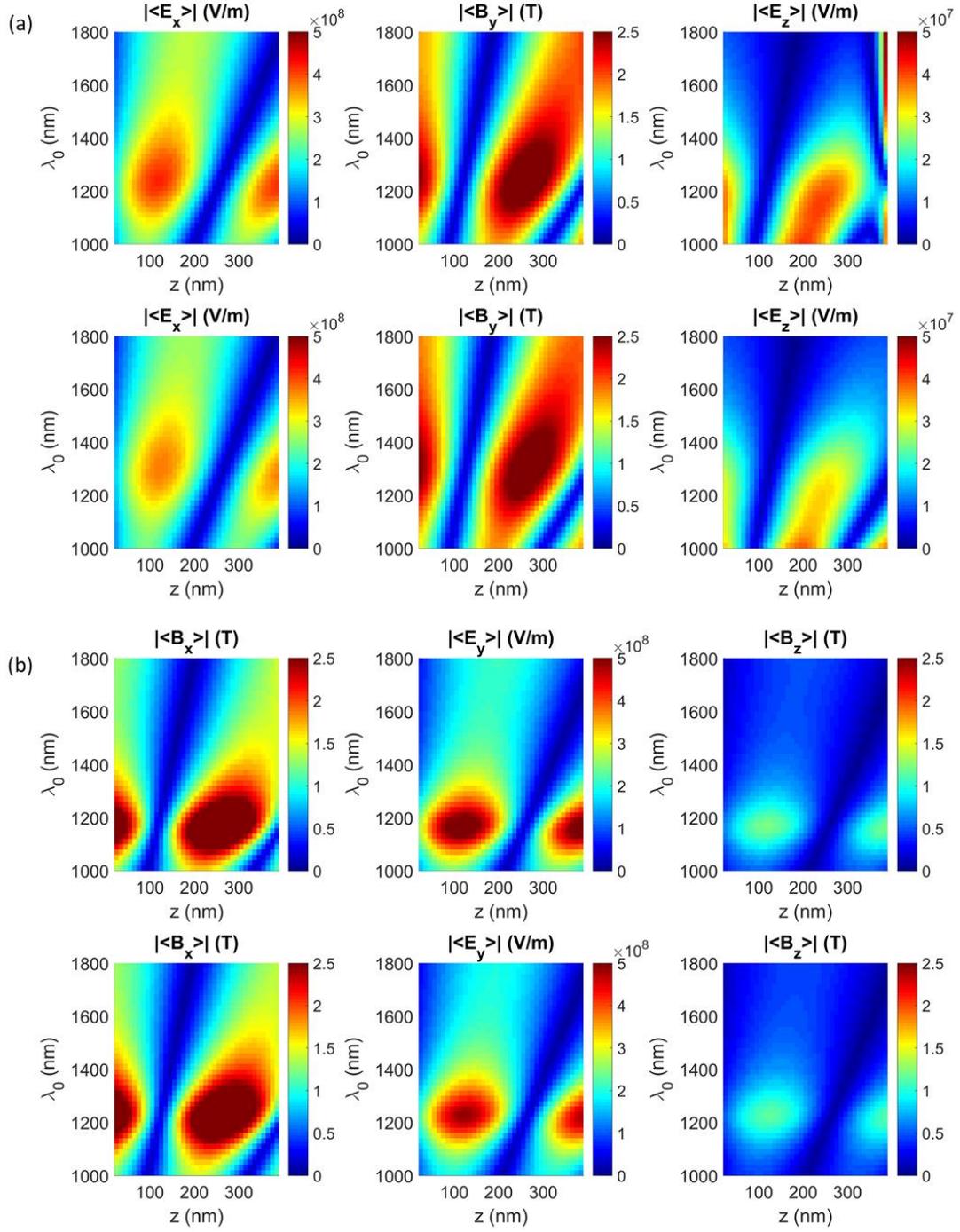

Fig. S7. Same as Fig. S5 for Sample C.

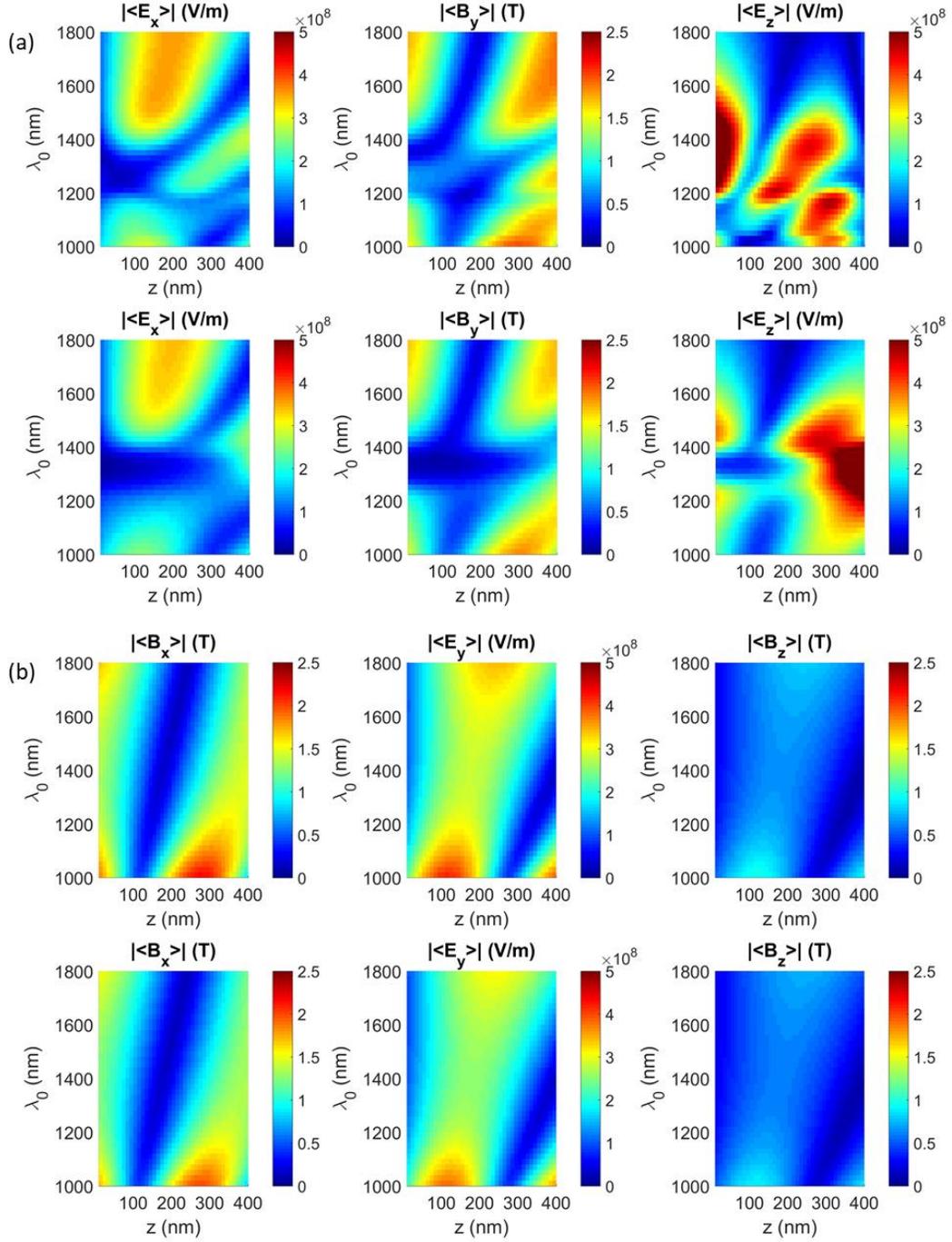

*Fig. S8. Same as Fig.S5 for Sample D. Note the disagreement between predictions of the EMT and the full-wave FEM calculations for p-polarized incident light in a wavelength range around 1400 nm corresponding to the elliptic and ENZ dispersion regimes of the metamaterial.*

## 5. Nonlinear response

The electromagnetic calculations described above yield spatial and spectral distributions of electromagnetic fields. According to the hydrodynamic model [40], the Maxwell's equations at the second harmonic frequency can be represented as:

$$\boldsymbol{\nabla} \times \boldsymbol{E^{(2\omega)}} = -\frac{\partial \boldsymbol{B^{(2\omega)}}}{\partial t}$$

$$\boldsymbol{\nabla} \times \boldsymbol{B^{(2\omega)}} = \frac{1}{c^2}\frac{\partial \boldsymbol{E^{(2\omega)}}}{\partial t} + \mu_0 \boldsymbol{j^{(2\omega)}}$$

$$\frac{\partial \boldsymbol{j^{(2\omega)}}}{\partial t} = -\gamma\,\boldsymbol{j^{(2\omega)}} + \frac{e^2 n_0}{m_e}\boldsymbol{E^{(2\omega)}} + \boldsymbol{S}$$

where nonlinear driving term $\boldsymbol{S}$ is given by

$$\boldsymbol{S} = \sum_\alpha \frac{\partial}{\partial r_\alpha}\left(\frac{\boldsymbol{j^{(\omega)}} j_\alpha^{(\omega)}}{en_0}\right) - \frac{e}{m_e}\left[\epsilon_0 (\boldsymbol{\nabla}\cdot\boldsymbol{E^{(\omega)}})\boldsymbol{E^{(\omega)}} + \boldsymbol{j^{(\omega)}}\times\boldsymbol{B^{(\omega)}}\right]$$

The equations above can be solved for the nonlinear current

$$\boldsymbol{j_{NL}^{(2\omega)}} = \frac{\boldsymbol{S}}{2\omega i + \gamma}$$

or, equivalently, for nonlinear polarization $\boldsymbol{P_{2\omega}}$ that is related to the nonlinear current via $\boldsymbol{j_{NL}^{(2\omega)}} = \frac{\partial \boldsymbol{P_{2\omega}}}{\partial t} = 2\omega i\,\boldsymbol{P_{2\omega}}$, resulting in Eq. (3) in the main text. The fields obtained from the FEM calculations can be directly used to derive the profile of the nonlinear driving term $\boldsymbol{S}$. Explicitly, in the full component form, the nonlinear source is given by

$$S_x = \frac{e}{m}\left(\varepsilon_0\left[E_x\frac{\partial E_x}{\partial x} + E_x\frac{\partial E_y}{\partial y} + E_x\frac{\partial E_z}{\partial z}\right] + j_y B_z - j_z B_y\right)$$
$$-\frac{1}{ne}\left(j_x\frac{\partial j_x}{\partial x} + j_y\frac{\partial j_x}{\partial y} + j_z\frac{\partial j_x}{\partial z} + \left[j_x\frac{\partial j_x}{\partial x} + j_x\frac{\partial j_y}{\partial y} + j_x\frac{\partial j_z}{\partial z}\right]\right)$$

$$S_y = \frac{e}{m}\left(E_y\varepsilon_0\left[\frac{\partial E_x}{\partial x} + \frac{\partial E_y}{\partial y} + \frac{\partial E_z}{\partial z}\right] + j_z B_x - j_x B_z\right)$$
$$-\frac{1}{ne}\left(j_x\frac{\partial j_y}{\partial x} + j_y\frac{\partial j_y}{\partial y} + j_z\frac{\partial j_y}{\partial z} + j_y\left[\frac{\partial j_x}{\partial x} + \frac{\partial j_y}{\partial y} + \frac{\partial j_z}{\partial z}\right]\right)$$

$$S_z = \frac{e}{m}\left(\varepsilon_0\left[E_z\frac{\partial E_x}{\partial x} + E_z\frac{\partial E_y}{\partial y} + E_z\frac{\partial E_z}{\partial z}\right] + j_x B_y - j_y B_x\right) - \frac{1}{ne}\left(j_x\frac{\partial j_z}{\partial x} + j_y\frac{\partial j_z}{\partial y} + j_z\frac{\partial j_z}{\partial z} + \right.$$
$$\left. \left[j_z\frac{\partial j_x}{\partial x} + j_z\frac{\partial j_y}{\partial y} + j_z\frac{\partial j_z}{\partial z}\right]\right) \qquad\qquad \text{(S1)}$$

We used the in-house codes to calculate derivatives that enter Eq. (S1) or, equivalently, Eq. (3) to avoid artifacts caused by the field discontinuities across the material boundaries and by non-uniform mesh inherent to commercial FEM implementations.

To validate the developed formalism, we first calculated second-harmonic generation spectra from a thin gold film (Fig. S9). As expected, the SHG response of the homogeneous metal film is dominated by $p \to p$ polarization configuration with a relatively weak $s \to p$ conversion and vanishing conversion of fundamental light to $s$ polarized SH. Importantly, the developed formalism predicts both spectral response as well as conversion ratio similar to that observed in the experiments.

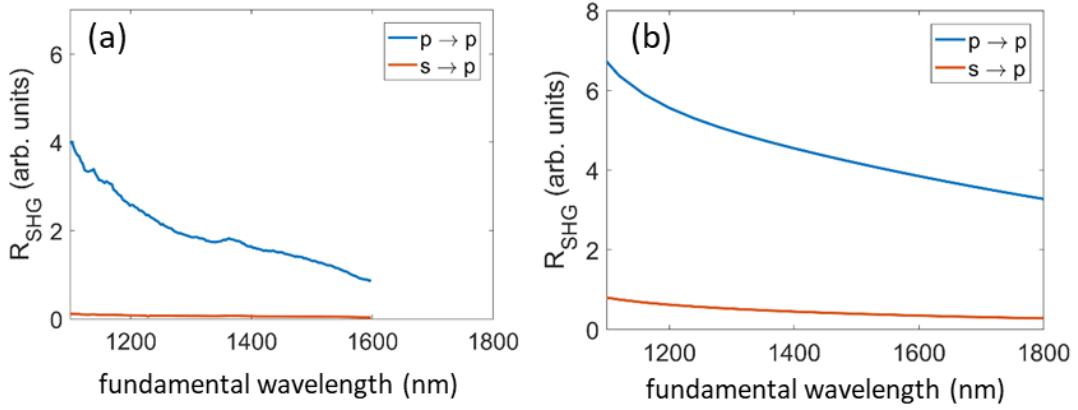

*Fig. S9. SHG response of a smooth Au film: (a) experiment and (b) numerical calculations.*

Having verified the numerical model on the example of a homogeneous metal film, we have turned to the calculations of SHG from the composites. In the limit of validity of local effective medium (that relies on quasi-static limit), the field is assumed to be homogeneous across the crossection of the nanowire. Furthermore, in the same limit, TM polarized light only has $E_x, E_z$, and $B_y$ field components while TE polarized waves have only $E_y, B_x, B_z$ field components.

The contributions to the source fields can be separated into two groups. The components of the first group produce the terms that do not vanish in the effective medium limit (when averaged over the unit cell):

$$S_x = \frac{e}{m}\left(-j_z B_y\right) - \frac{1}{ne}\left(j_z \frac{\partial j_x}{\partial z} + j_x \frac{\partial j_z}{\partial z}\right)$$

$$S_z = \frac{e}{m}\left(j_x B_y\right) - \frac{1}{ne}\left(j_z \frac{\partial j_z}{\partial z} + j_z \frac{\partial j_z}{\partial z}\right) \qquad \text{(S2a)}$$

for the TM polarized excitation and

$$S_x = \frac{e}{m}\left(j_y B_z\right)$$

$$S_z = \frac{e}{m}\left(-j_y B_x\right) \qquad \text{(S2b)}$$

for the TE-polarized excitation. The second group contains the remaining terms in Eq. (S1).

Starting from Eq.(3), it becomes possible to use the Maxwell-Garnett EMT to convert the nonlinear source terms in a quadratic form [Eq. (4)]. In doing so, we utilize the relationship between the components of the electric and magnetic fields in the plane wave, $k_x H_y = -\frac{\omega}{c} D_z$. The spectral and spatial distributions of the unit-cell averaged nonlinear sources, calculated by the full-wave solutions of the Maxwell's equations and predicted by the nonlinear EMT are shown in Fig. S10—S13. The nonlinear EMT adequately describes these distributions with the exception of the red-shift similar to what was observed in the linear reflection spectra (Fig. S4). Apart from the red-shift, the disagreement between the nonlinear EMT and the full-wave solutions is limited to proximity to the wire edges and to the low-frequency range in Sample D that can be attributed to the excitation of an additional wave in nanowire composites [30].

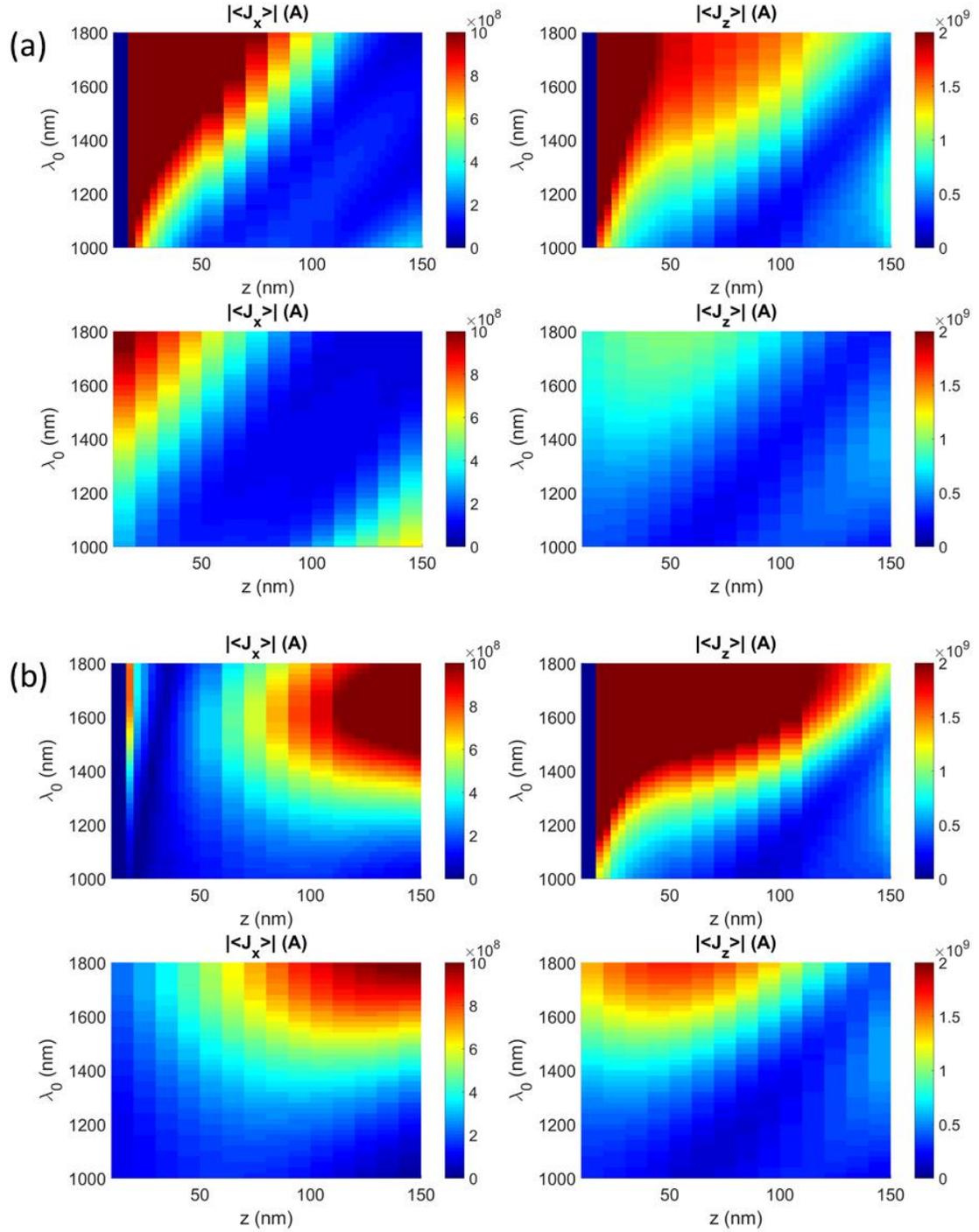

Fig. S10. Spatial and spectral distributions of the unit-cell-averaged non-vanishing components of the nonlinear current calculated with the FEM (top) and the nonlinear EMT (bottom) for Sample A excited with p-polarized (a) and s-polarized (b) light.

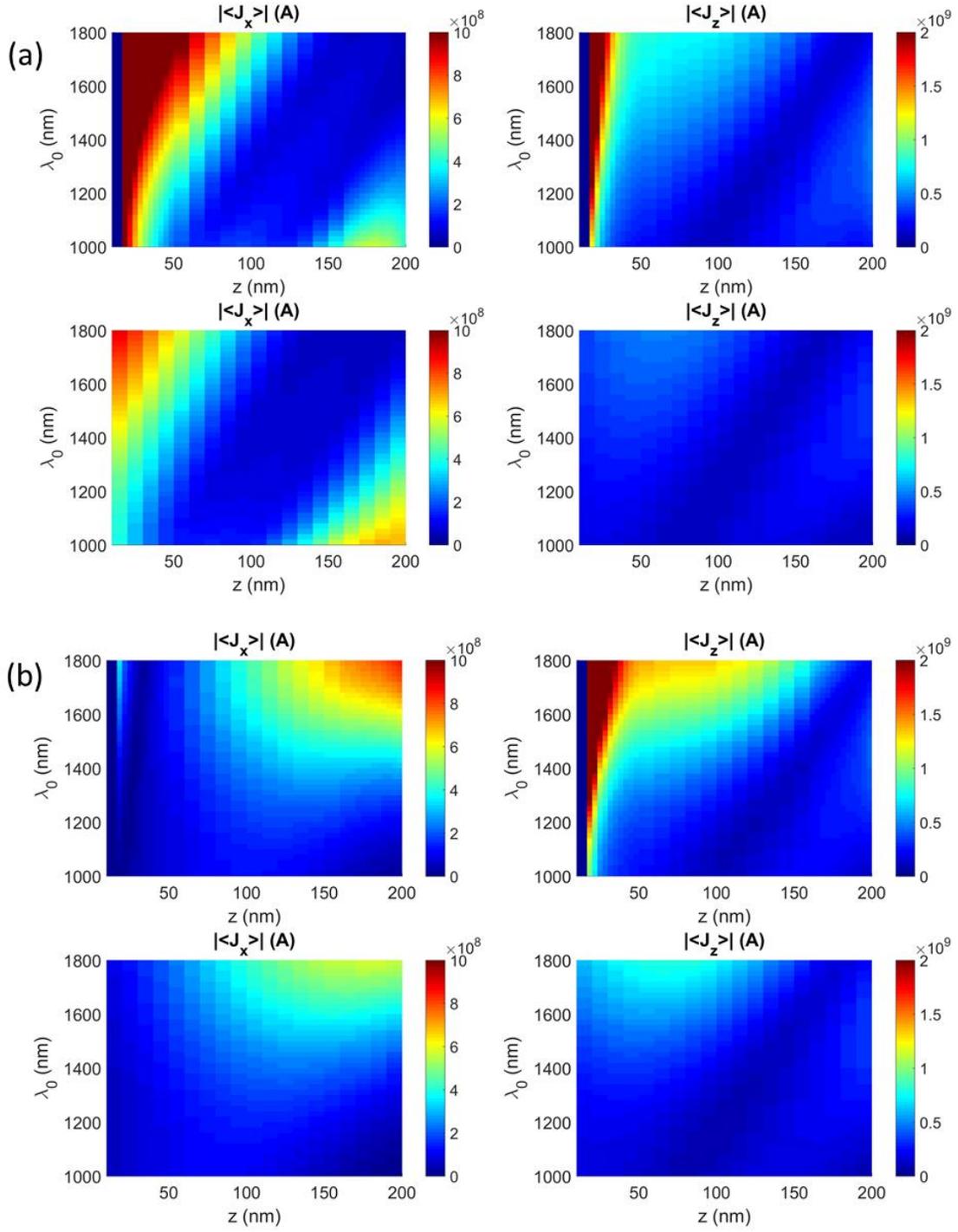

*Fig. S11. Same as Fig. S10 for Sample B.*

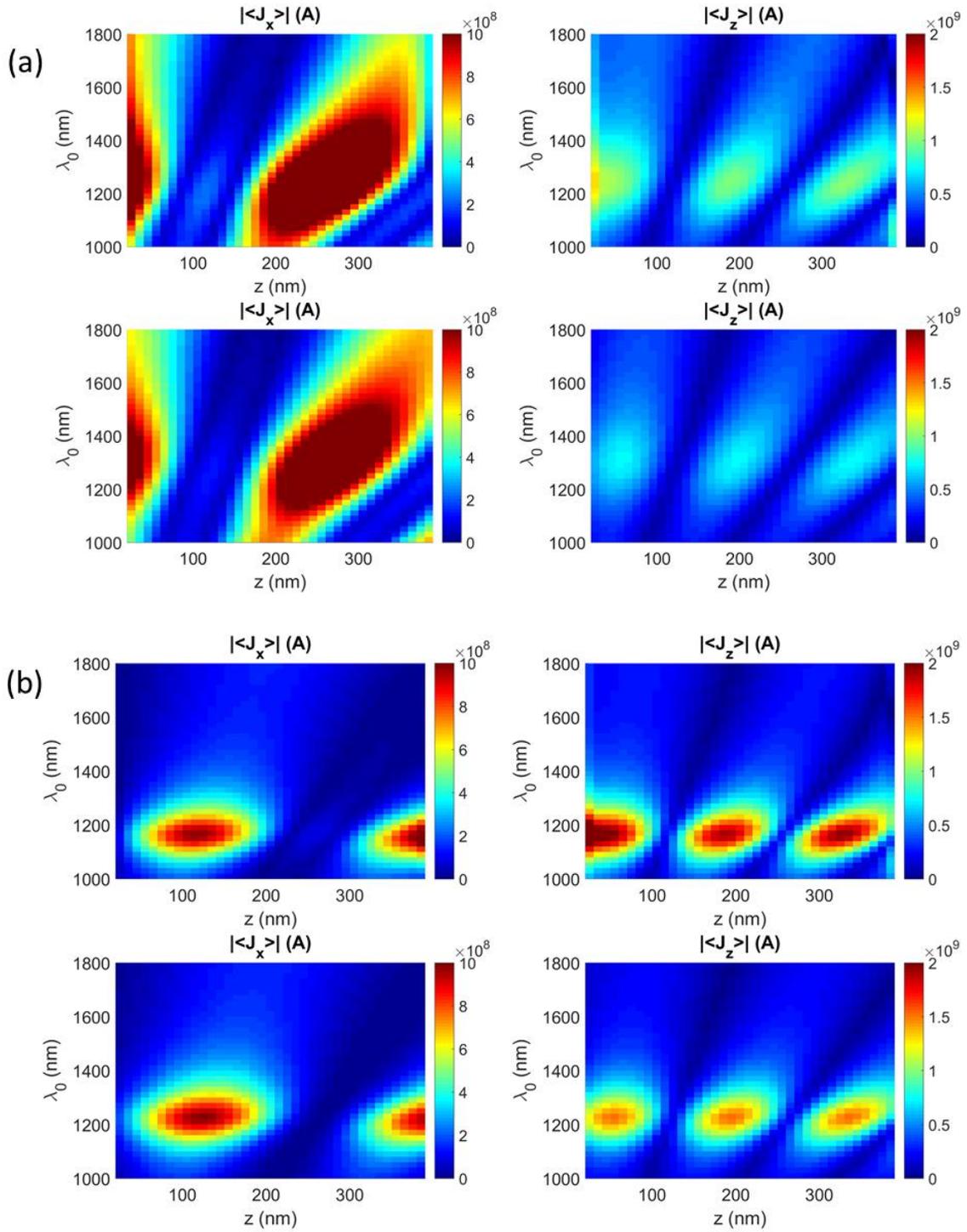

Fig. S12. Same as Fig. S10 for Sample C.

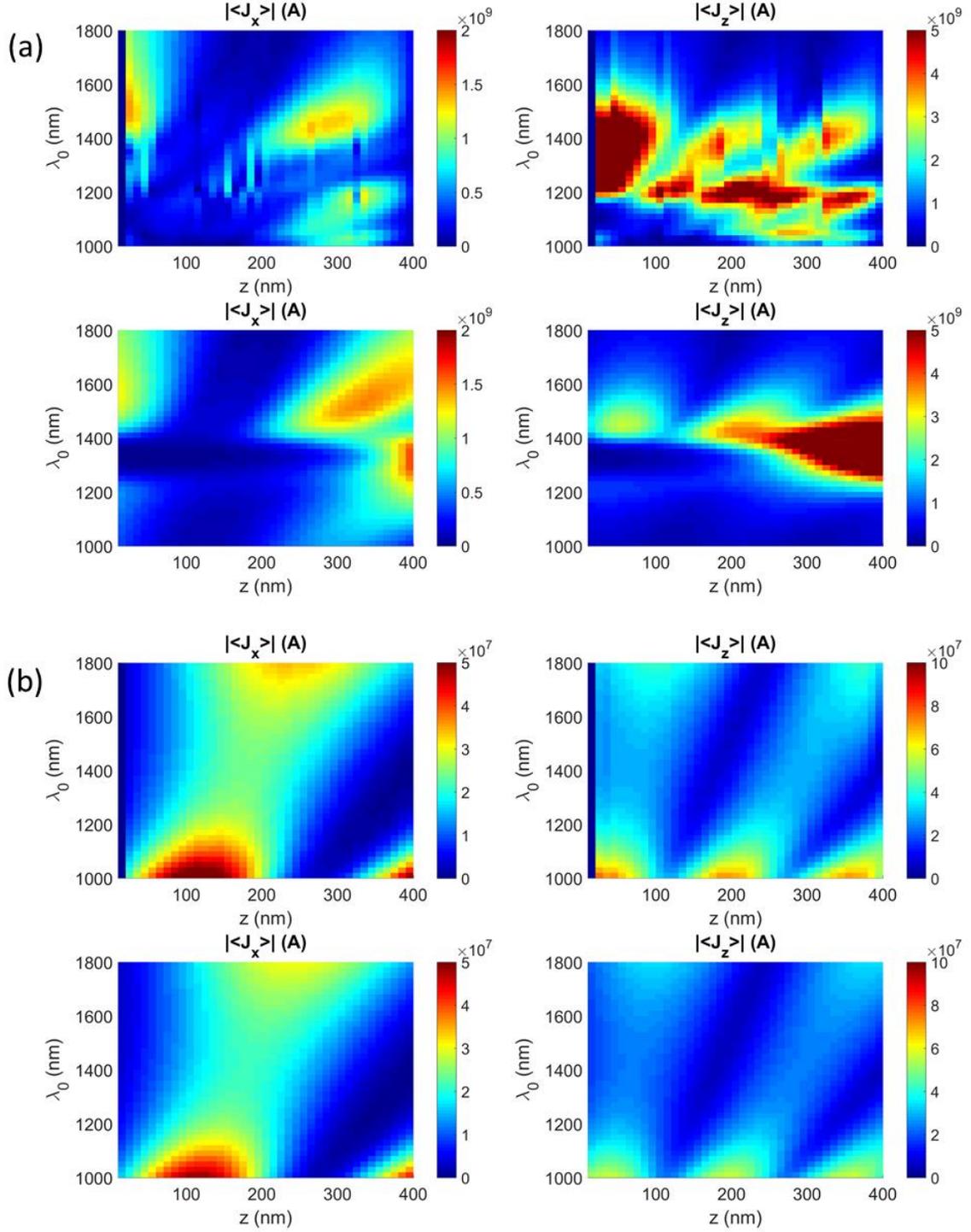

*Fig. S13. Same as Fig. S10 for Sample D.*

To identify the terms that dominate SHG emission in metamaterials, we have run the full-wave 3D calculations using the complete distributions of the nonlinear current sources as well as the nonlinear current sources attributed to the two sub-groups of the terms of Eq. (S1) as described above. The results of these calculations are summarized in Figs. 2 ,4, S14, and S15. It is clearly seen that the SHG is dominated by the terms included in Eq. (S2) that are well described by the nonlinear EMT.

Finally, in order to test the limits of validity of the nonlinear EMT, we have incorporated this model into quasi-2D calculations that assume that the metamaterial can be homogenized at both fundamental and second harmonic frequencies. The results of these calculations are shown in insets in Figs. S14d and S15d. It is seen that treating the composite as an effective medium at the second harmonic frequency results in qualitatively and quantitatively wrong predictions of second harmonic generation. We attribute the difference between predictions of the fully-homogenized model and full 3D treatment of SHG generation to substantial coupling of second harmonic light to cylindrical surface plasmon modes supported by the wires. Propagation of these modes is not adequately described by the local effective medium theory but it can be included in the effective medium description by introducing nonlocal effective permittivity [30]. It is, therefore, possible that further development of the second harmonic EMT presented in this work in order to incorporate more rigorous homogenization of the composite, may result in adequate fully-homogenized description of the second harmonic nonlinearity in the nanowire metamaterials.

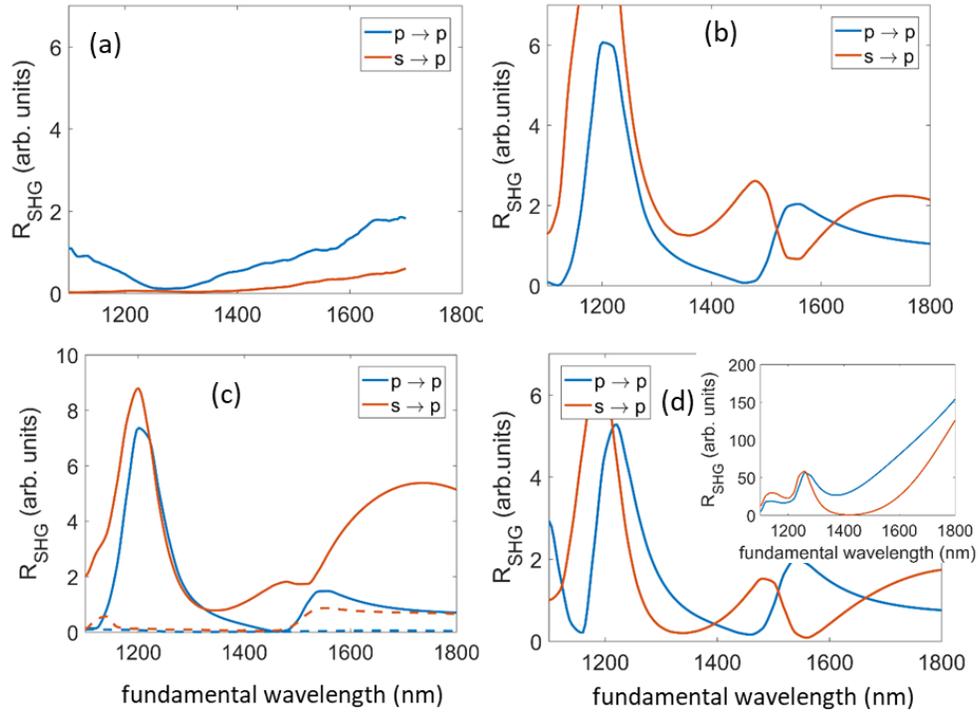

Fig. S14. Spectra of second harmonic generation from Sample B: (a) experiment, (b) the full-wave numerical solutions of the Maxwell's equations, (c) is similar to (b) with the SHG contribution from (solid lines) the terms included in Eq. (S2) and (dashed lines) the remaining terms in Eq. (S1) considered separately, and (d) the results of the nonlinear EMT with the SHG intnesity calculated with the full 3D model (main figure) and the EMT (inset).

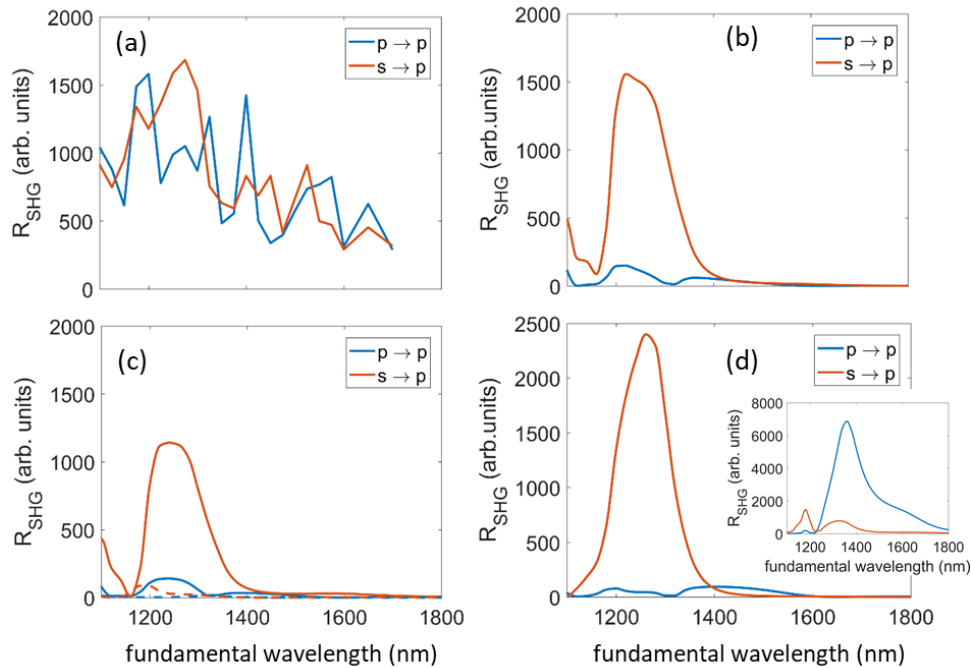

*Fig. S15. Same as Fig. S14 for Sample C. In the experiment, same parameters described in the main text were used, apart from a beam spot size of 4 μm.*


\end{document}